\documentstyle[12pt]{article}

\newcommand{\be}{\begin{equation}}
\newcommand{\ee}{\end{equation}}
\newcommand{\beq}{\begin{eqnarray}}
\newcommand{\eeq}{\end{eqnarray}}

\newcommand{\eps}{\varepsilon}

\newcommand{\tE}{\lefteqn{\smash{\mathop{\vphantom{<}}\limits^{\;\sim}}}E}
\newcommand{\tP}{\lefteqn{\smash{\mathop{\vphantom{<}}\limits^{\;\sim}}}P}
\newcommand{\tQ}{\lefteqn{\smash{\mathop{\vphantom{<}}\limits^{\;\sim}}}Q}
\newcommand{\Et}{\lefteqn{\smash{\mathop{\vphantom{\Bigl(}}\limits_{\sim}
\atop \ }}E}
\newcommand{\Pt}{\lefteqn{\smash{\mathop{\vphantom{\Bigl(}}\limits_{\sim}
\atop \ }}P}
\newcommand{\Qt}{\lefteqn{\smash{\mathop{\vphantom{\Bigl(}}\limits_{\sim}
\atop \ }}Q}

\newcommand{\tN}{\lefteqn{\mathop{\vphantom{\Bigl(}}\limits_{\sim}
\atop \ }N}
\newcommand{\tM}{\lefteqn{\mathop{\vphantom{\Bigl(}}\limits_{\,\sim}
\atop \ }M}
\newcommand{\tNn}{\lefteqn{\mathop{\vphantom{\Bigl(}}\limits_{\sim}
\atop \ }{\cal N}}

\newcommand{\tPb}{{\tP_{\smash{(\beta)}}}}
\newcommand{\nd}{{\cal N}_D}

\newcommand{\ba}{\beta_1}
\newcommand{\bb}{\beta_2}
\newcommand{\F}{{\cal F}}
\newcommand{\G}{{\cal G}}
\newcommand{\FF}{\Phi}

\newcommand{\sqr}[1]{\left( #1 \right)^{1/2}}

\begin{document}
\title{ \Large \bf
SO(4,C)-covariant Ashtekar--Barbero gravity and the Immirzi parameter}
\author{\normalsize S.~Alexandrov\thanks{e.mail: alexand@snoopy.phys.spbu.ru}
\\
{ \small \it Department of Theoretical Physics,
St.Petersburg University,198904 St.Petersburg, Russia}
}
\date{}
\maketitle

\begin{abstract}
An so(4,C)-covariant hamiltonian formulation of a family of generalized
Hilbert--Palatini actions depending on a parameter (the so called Immirzi
parameter) is developed.
It encompasses the Ashtekar--Barbero gravity which serves as a basis of
quantum loop gravity.
Dirac quantization of this system is constructed.
Next we study dependence of
the quantum system on the Immirzi parameter. The path integral quantization
shows no dependence on it. A way to modify the loop approach
in the accordance with the formalism developed here is briefly outlined.
\end{abstract}

\section{Introduction}

The construction of the complete theory of quantum gravity is still
an open problem. There are several approaches to quantization
of general relativity and to understanding what the quantum spacetime is.
One of the most promising approaches is loop quantum gravity
\cite{loop}, \cite{cyl}, \cite{Rov}.
It is mathematically well-defined and explicitly background independent.
In this framework a set of remarkable physical results has been obtained
such as the discrete spectra of the area and volume operators
\cite{area}, \cite{volume}
and a derivation of the Bekenstein--Hawking formula for
the black hole entropy \cite{entropy}.

However, there are still several important problems.
One of them is the so called Immirzi parameter problem.
This problem arises due to the results obtained for the spectra of
the geometrical operators and the black hole entropy are
proportional to an unphysical parameter. It is called "Immirzi parameter"
\cite{Rov-Tim} and appears as a parameter of a canonical transformation
in classical gravity \cite{barb}, \cite{imir}.
So at least at the classical level this parameter should be unphysical.
The problem is whether the quantum theory can nevertheless depend on it
and, if not, why we observe this dependence for the physical quantities.
While this problem is not resolved, it does not allow to
interpret the discrete spectra of the area and volume
as evidence for a discrete structure of
spacetime. Several interpretations
\cite{Rov-Tim} of this dependence have been proposed
but there is no any acceptable explanation yet.

In this paper we suggest a new strategy to tackle
the Immirzi parameter problem. It is based on the use of a larger
symmetry group
\footnote{The idea that the problem of the Immirzi parameter
requires a group larger than SU(2) has been stressed by Immirzi himself
in the work where the problem has been discovered \cite{imir}.}.
Namely, our aim is to develop the canonical formalism for
gravity with the full Lorentz gauge group in the tangent space.
In contrast to the standard approach we do not impose
any gauge fixing like "time gauge" used to obtain the Ashtekar--Barbero
formulation \cite{barb}, which lies in the ground of quantum loop gravity.
However, we are still able to use a connection as a canonical variable,
but it turns out to be so(4,C) connection rather than su(2).
It is possible since the formalism can be put in an so(4,C) covariant form
so that all calculations are carried out in a nice and elegant way.

Then this covariant representation is used for
the investigation of dependence
of the quantum theory on the Immirzi parameter.
In this paper we give only
some preliminary considerations
in the frameworks of the path integral quantization and
a modified loop approach.
They can gain an insight on
the Immirzi parameter problem. However,
more elaborated technics is needed to set these considerations
on a solid ground.

The paper is organized as follows.
In the next section the 3+1 decomposition of the generalized
Hilbert--Palatini action is obtained using the results from Ashtekar
gravity \cite{Ash-1}, \cite{Abook}.
The decomposed action is presented in an so(4,C) covariant form.
In section 3 the hamiltonian formalism is constructed and the canonical
commutation relations are obtained.
Section 4 is devoted to application of the developed formalism
to the investigation of dependence of the quantum theory on
the Immirzi parameter.
In the first subsection the path integral for gravity described by
the generalized Hilbert--Palatini action is shown to be
independent on the Immirzi parameter in the so called Yang--Mills gauge.
The second one is intended to present
some ideas how this formalism can be put in the ground of the loop approach
and how it can cure the Immirzi parameter problem.
Some concluding remarks are placed in section 5.
The appendices contain some general formulas and examples.

We use the following notations for indices.
The indices
$i,j,\dots$ from the middle of the alphabet label the space coordinates.
The latin indices $a,b,\dots$ from the beginning of the alphabet
are the su(2) indices, whereas
the capital letters $X,Y,\dots$ from the end of the alphabet are
the so(3,1) or so(4,C) indices.

\section{Generalized Hilbert--Palatini action}

We start with the generalized Hilbert--Palatini action suggested by Holst
\be
S_{(\beta)}=\frac12 \int \eps_{\alpha\beta\gamma\delta}
e^\alpha \wedge e^\beta \wedge (\Omega^{\gamma\delta}+
\frac{1}{\beta}\star\Omega^{\gamma\delta}). \label{Sd1}
\ee
Here the star operator is defined as
$ \star \omega^{\alpha\beta}=
\frac 12 {\eps^{\alpha\beta}}_{\gamma\delta}
\omega^{\gamma\delta} $ and $\Omega^{\alpha\beta}$ is the curvature of
the spin-connection $\omega^{\alpha\beta}$.
In the work \cite{Holst} Holst has shown that in the "time" gauge
this action reproduces Barbero's formulation \cite{barb} and
the parameter $\beta$ plays the role of the Immirzi parameter.
Our aim is to investigate the action (\ref{Sd1}) without imposing
any gauge fixing. As we shall see the hamiltonian formulation of the theory
described by this action allows a remarkable covariant representation.

To this end, let us do the 3+1 decomposition. It easily can be obtained
from the decomposition of the self-dual Hilbert--Palatini action ($\beta=i$)
leading to Ashtekar gravity.
Such decomposition in suitable variables without a gauge fixing
has been obtained in \cite{AV} and looks like
\beq
S_A&=&2\int dt\ d^3x (P^i_{(i)a}\partial_tA_i^{(i)a}+
A_0^{(i)a}{\cal G}^{(A)}_a+{\nd}^i H^{(A)}_i+{\tNn} H^{(A)}), \label{Asht} \\
{\cal G}^{(A)}_a&=&\nabla_iP_{(i)a}^i=\partial_iP_{(i)a}^i
-{\eps_{ab}}^cA^{(i)b}_iP^{i}_{(i)c} , \nonumber \\
 H^{(A)}_i&=&-P^k_{(i)a}F^{(i)a}_{ik},
\nonumber \\
H^{(A)}&=&-\frac12 P^i_{(i)a}P^j_{(i)b} {\eps^{ab}}_c F^{(i)c}_{ij},
\nonumber \eeq
where
\beq
F^{(i)a}_{ij}&=&\partial_i A_j^{(i)a} -\partial_j A_i^{(i)a}-
{\eps^a}_{bc} A_i^{(i)b} A_j^{(i)c}, \nonumber \\
A_i^{(i)a}&=&\xi^a_i-i\zeta^a_i=\frac12\,{\eps^a}_{bc}\omega_i^{bc}
-i\omega_i^{0a},  \\
P^i_{(i)a}&=&{\eps_a}^{bc}\tE^i_b\chi_c+i\tE^i_a. \nonumber
\eeq
The label $(i)$ refers to the value of $\beta$.
The Lagrange multipliers $\nd^i$, $\tNn$, triad $\tE^i_a$ and field $\chi_a$
arise from the decomposition of the tetrad:
\beq
&e^0=Ndt+\chi_a E_i^a dx^i ,\quad e^a=E^a_idx^i+E^a_iN^idt,& \nonumber \\
&{\tE}^i_a =h^{1/2}E^i_a,  \quad
\tN=h^{-1/2} N, \quad \sqrt{h}=\det E^a_i,& \label{tetrad} \\
&N^i=\nd^i+\tE^i_a\chi^a\tNn, \quad
\tN=\tNn+\Et_i^a\chi_a\nd^i.& \nonumber
\eeq
Here $E^i_a$ is the inverse of $E_i^a$.
The field $\chi_a$ describes deviation
of the normal to the spacelike hypersurface $\{ t=0\}$ from
the time direction.

Let us return to the generalized Hilbert--Palatini action (\ref{Sd1}).
It is related to the Ashtekar action by
\be S_{(\beta)}={\rm Re} S_A-\frac{1}{\beta}{\rm Im} S_A. \label{Sd2}\ee

It is convenient to combine pairs of the su(2) indices $a$ into
a single so(3,1) index $X$. Accordingly we can introduce
the so(3,1) multiplets:
 \beq
 & \G_X=({\cal L}_a,{\cal G}_a)=
 ({\rm Im}{\cal G}^{(A)}_a, {\rm Re}{\cal G}^{(A)}_a)
     &{\rm -\ constraint\ multiplet}, \nonumber\\
 & A_i^{ X}=(\zeta^a_i,\xi_i^a)
     &{\rm -\ connection\ multiplet}, \nonumber\\
&  \tP_X^{ i}=(\tE^i_a,{\eps_a}^{bc}\tE^i_b\chi_c)
     &{\rm -\ first\ triad\ multiplet}, \label{multHP}\\
&  \tQ_X^{ i}=(-{\eps_a}^{bc}\tE^i_b\chi_c,\tE^i_a)
     &{\rm -\ second\ triad\ multiplet}, \nonumber\\
&  \tPb_X^i=\tP_X^i-\frac{1}{\beta}\tQ_X^i
     &{\rm -\ dynamical\ triad\ multiplet}.  \nonumber
 \eeq
The relations between the triad multiplets are presented in Appendix A.
As it will be shown the fields form multiplets in
the adjoint representation
of so(4,C) with the following structure constants
\footnote{
Indeed the given representation corresponds to the so(3,1) algebra, which is
a real form of so(4,C). However, we are free to do complex gauge
transformations as well as complex transformations of the basis
of the adjoint representation (see below). An example,
which involves imaginary structure constants and fields explicitly, is
presented in Appendix B.
So the gauge algebra allows its extension to so(4,C).
}:
\be
\begin{array}{ccc}
f_{A_1 A_2}^{A_3}=0,&
f_{A_1 B_2}^{A_3}=-\eps^{A_1 B_2 A_3},&
f_{B_1 B_2}^{A_3}=0, \\
f_{B_1 B_2}^{B_3}=-\eps^{B_1 B_2 B_3},&
f_{A_1 B_2}^{B_3}=0,&
f_{A_1 A_2}^{B_3}=\eps^{A_1 A_2 B_3}.
\end{array}      \label{algHP}
\ee
Here we split the 6-dimensional index $X$ into a pair of 3-demensional
indices, $X=(A,B)$, so that $A,B=1,2,3$. $\eps$ is the Levi--Civita
symbol, $\eps^{123}=1$.

With these multiplets using (\ref{Sd2}) and (\ref{Asht}) the action
(\ref{Sd1}) can be represented in the form:
\beq
S_{(\beta)} &=&\int dt\, d^3 x (\tPb^i_X\partial A^X_i
+{\cal N}_{\G}^X \G_X+\nd^i H_i+\tNn H),  \label{Sd3} \\
\G_X&=&\partial_i \tPb^i_X +f_{XY}^Z A^Y_i \tPb^i_Z,\nonumber \\
H_i&=&-\tPb^j_X F_{ij}^X , \nonumber \\
H&=&-\frac12 \tPb^i_X \tP^j_Y f^{XY}_Z F_{ij}^Z, \nonumber \\
F^X_{ij}&=&\partial_i A_j^X-
\partial_j A_i^X+f_{YZ}^X A^Y_i A^Z_j, \nonumber
\eeq
where we have used the Killing form to raise and low indices
$f^{XY}_Z=g^{XX'}g^{YY'}g_{ZZ'}f^Z_{XY}$:
\be
g_{XY}=f_{XZ_1}^{Z_2}f_{YZ_2}^{Z_1},\quad
g^{XY}=(g^{-1})^{XY}, \quad
g_{XY} =\left(
\begin{array}{cc}
\delta_{ab}&0 \\ 0&-\delta_{ab}
\end{array}
\right).
\ee

As a result we have represented the 3+1 decomposed action in the so(4,C)
covariant form.
Moreover, it is covariant under arbitrary transformations
of the basis of the adjoint representation. If we change
\be \G_X \longrightarrow (U^{-1})_X^Y \G_Y, \quad
  \tE_X^i \longrightarrow (U^{-1})_X^Y \tE^i_Y, \quad
  A_i^X \longrightarrow U^X_Y A_i^Y,
\ee
where $U^X_Y$ is an arbitrary invertable matrix and
$\tE^i_X$ denotes any triad multiplet, the representation
(\ref{Sd3}) is unchanged.
An example of a formulation in such rotated basis, which appears naturally
for the action (\ref{Sd1}) rewritten in terms of curvatures only without
the star operator, is presented in Appendix B.
From now on
$\G_X$, $\tE^i_X$, $A_i^X$ will
denote multiplets in an arbitrary basis.

\section{Covariant canonical formulation}

Let us construct the canonical formalism for the
action (\ref{Sd3}).
From the beginning $A_i^X$ and $\tPb^i_X$ are canonical variables, i.e.,
\beq \{A_i^X,\tPb_Y^j\}&=&\delta^j_i \delta^X_Y, \nonumber \\
     \{A_i^X,\tQ_Y^j\}&=&\delta^j_i (\Lambda^{-1})^X_Y. \eeq
However, there are constraints on the momenta. In the covariant form
they can be represented as
\be \phi^{ij}=\Pi^{XY}\tQ^i_X\tQ^j_Y. \label{phi} \ee
The matrices $\Pi$ and $\Lambda$ appearing in the formulas above
are introduced in Appendix A (\ref{p-q}), (\ref{pb-q}).
$\phi^{ij}$ is symmetric, so there are only six independent constraints.
It is clear that $\{ \G_X,\phi^{ij}\}=0$ and $\{ H_k ,\phi^{ij}\}\sim
\phi^{ij}$. The only nontrivial bracket is with
the hamiltonian constraint. Using (\ref{Sd3}), (\ref{inverse}),
(\ref{com-f}), (\ref{ff}) we obtain
\beq
\{ H,\phi^{ij}\}&=& 2\tQ^n_X\tQ^{\{ j}_Y
f^{XYZ}\left( \partial_{n}\tQ_Z^{i\}}+f^{T}_{ZS} A_{n}^S\tQ_T^{i\}}  \right)
\nonumber \\
&=& \psi^{ij}+2\phi^{i[j}\tP^{n]}_Z A^Z_n,
\label{psi-phi}
\eeq
where
\beq
\psi^{ij}&=&2f^{XYZ}\tQ_X^{n}\tQ_Y^{\{ j}\partial_n \tQ_Z^{i\} }
-2(\tQ\tQ)^{\{ i[j\} }\tQ_Z^{n]}A_n^Z, \label{psi} \\
(\tQ\tQ)^{ij}&=&g^{XY}\tQ_X^i\tQ_Y^j
\eeq
and symmetrization is taken with the weight $1/2$,
whereas antisymmetrization does not include any weight.
It is remarkable that the second class constraints (\ref{phi}) and
(\ref{psi}) don't depend on $\beta$ that proves consistency of
the constraints in different formulations.

One can calculate\footnote{If we defined the secondary second class constraint as
the full r.h.s. of (\ref{psi-phi}), the Poisson bracket $D_2$ would vanish whereas
$D_1$ would remain the same. This could be advantageous for calculation of the Dirac
bracket defined below.}
\beq
D_1^{(ij)(kl)}=
\{ \phi^{ij}, \psi^{kl}\}&=&\frac{4\beta^2}{1+\beta^2}(\tQ\tQ)^{\{i[j\}}
(\tQ\tQ)^{\{k]l\}} , \label{D1} \\
D_2^{(ij)(kl)}=
\{ \psi^{ij}, \psi^{kl}\} &\approx& \frac{8\beta^2}{1+\beta^2}\left[
(\tQ\tQ)^{\{i\{k}f^{XYZ}\tQ^{l\}}_X \tQ^{j\}}_Y\partial_n \tP_Z^{n}
\right.
\nonumber \\
&& \left. + (\tQ\tQ)^{ij}f^{XYZ}\tP^n_X\tQ^{\{ k}_Y \partial_n \tQ_Z^{l\}}
-(\tQ\tQ)^{kl}f^{XYZ}\tP^n_X\tQ^{\{ i}_Y \partial_n \tQ_Z^{j\}}
\right], \label{D2}
\eeq
where in the second equality we used the Gauss constraint $\G_X$
and both second class constraints (\ref{phi}) and
(\ref{psi}).

Let us redefine the constraints $\FF_{\alpha}=(\G_X,H_i,H)$:
\be {\tilde \FF}_{\alpha}=\FF_{\alpha}-\phi^{ij}(D_1^{-1})_{(ij)(kl)}
\{ \FF_{\alpha},\psi^{kl}\}. \ee
Then ${\tilde \FF}_{\alpha}$ are first class constraints with the
algebra presented in Appendix C.
 The remaining constraints are second class.
 They form the matrix of commutators:
\be \Delta=\left(
\begin{array}{cc}
0 & D_1 \\
-D_1 & D_2
\end{array} \right), \quad
\Delta^{-1}=\left(
\begin{array}{cc}
D_1^{-1} D_2 D_1^{-1} & -D_1^{-1} \\
D_1^{-1} & 0
\end{array}  \right).   \label{Delta}
\ee
It gives rise to the Dirac bracket \cite{Dirac}
\be \{ K,L\}_D =\{ K,L\}-\{ K,\varphi_{r}\}(\Delta^{-1})^{rr'}
\{\varphi_{r'},L\}, \ee
where $\varphi_r=(\phi^{ij},\psi^{ij})$.
However, it is simplified when one of the functions coincides with
the first class constraint $\FF_{\alpha}$ (or ${\tilde \FF}_{\alpha}$).
Then
\be \{ \FF_{\alpha},L\}_D \approx \{ \FF_{\alpha},L\}-
\{ \FF_{\alpha},\psi^{ij}\}(D_1^{-1})_{(ij)(kl)} \{\phi^{kl},L\}.
\label{D-con} \ee
From (\ref{psi-phi}) and the Jacoby identity one can see that
$\{ \G_X,\psi^{ij}\}=0$ and $\{ H_k ,\psi^{ij}\}\sim \psi^{ij}$.
Thus the last term in (\ref{D-con}) survives only in the case
when $\FF_{\alpha}$ is
the hamiltonian constraint and $L$ depends on the connection variables.
In all other cases the Dirac bracket coincides with the ordinary one.

This fact is the reason for the remarkable relations between
the brackets in different formulations, which can be easily
checked by a direct calculation:
\beq
\{ \FF_{\alpha}^{(\beta)},\tPb^i_X\}_D^{(\beta)}&=&
\{ \FF_{\alpha}^{(\beta')},\tPb^i_X\}_D^{(\beta')}, \nonumber \\
\{ \FF_{\mu}^{(\beta)},A_i^X\}_D^{(\beta)}&=&
\{ \FF_{\mu}^{(\beta')},A_i^X\}_D^{(\beta')}. \label{remark}
\eeq
Here the label $(\beta)$ indicates the formulation which a given object
are taken from. The last equality is not valid for the hamiltonian
constraint so that $\FF_{\mu}=(\G_X,H_i)$.

Using the coincidence of the Dirac and Poisson brackets (\ref{D-con})
the transformation laws of the multiplets can easily be found:
\beq
\{ \G_X,\G_Y\}_D&=&f_{XY}^Z \G_Z, \nonumber \\
\{ \G_X,A^Y_i\}_D&=&
\delta^Y_X\partial_i -f_{XZ}^Y A^Z_i,  \label{trans} \\
\{ \G_X,\tE_Y^i\}_D&=&f_{XY}^Z \tE_Z^i.  \nonumber
\eeq
As it was declared they form the adjoint representation of so(4,C).
$A_i^X$ is a true so(4,C)-connection.

To be sure that we do have the so(4,C) gauge algebra rather than
its real form only, one should check that complex gauge transformations
are allowed. The criterion is reality of the 3-dimensional metric and
its evolution:
\be
 g^{ij}=g^{XY}\tP^i_X\tP^j_Y=
\frac{\beta^2}{1+\beta^2}\,g^{XY}\tPb^i_X\tPb^j_Y. \label{3-met}
\ee
Since $\{ \G_X, g^{ij}\}_D=0$, the complex gauge transformations do not
destroy reality of the metric. Similarly, the time evolution remains
real due to
\beq &\{ \G_X, \frac{d}{dt}g^{ij} \}_D=\{\G_X,
\{ \int d^3 x\,({\cal N}_{\G}^Y \G_Y+\nd^i H_i+\tNn H), g^{ij}\}_D \}_D &
\nonumber \\
&= \{ \{ \G(n),\int d^3 x\,({\cal N}_{\G}^X \G_X+\nd^i H_i+\tNn H)\}_D,
g^{ij}\}_D =0.&
\eeq

To complete the construction of the canonical formalism, we should
find $D_1^{-1}$. To this end, introduce the inverse triad multiplets:
\beq
\Pt_i^{X}&=&\left( \frac{\delta^a_b-\chi^a\chi_b}{1-\chi^2}\Et_i^b,
-\frac{ {\eps^a}_{bc}\Et_i^b\chi^c}{1-\chi^2} \right),
\nonumber \\
\Qt_i^{X}&=&\left( \frac{ {\eps^a}_{bc}\Et_i^b\chi^c}{1-\chi^2},
\frac{\delta^a_b-\chi^a\chi_b}{1-\chi^2}\Et_i^b \right).
\label{invQ}\eeq
They obey the following equations:
\beq
& \{ \G_X,\Qt_i^Y\}=-f_{XZ}^Y\Qt_i^Z,\quad
\{ \G_X,\Pt_i^Y\}=-f_{XZ}^Y\Pt_i^Z,& \nonumber \\
& \tQ^i_X\Qt_j^X=\delta^i_j.  \quad
\tP^i_X\Pt_j^X=\delta^i_j, & \\
& \tQ^i_X\Pt_j^X=\tP^i_X\Qt_j^X=0, & \nonumber
\eeq
In the basis (\ref{multHP}) one can obtain
\beq
 I_{(P)X}^Y &\equiv& \tP^{ i}_X\Pt_i^{Y}=\left(
\begin{array}{cc}
\frac{\delta_a^b-\chi_a\chi^b}{1-\chi^2} &
\frac{ {\eps_a}^{bc}\chi_c}{1-\chi^2} \\
\frac{ {\eps_a}^{bc}\chi_c}{1-\chi^2} &
-\frac{\delta_a^b\chi^2-\chi_a\chi^b}{1-\chi^2}
\end{array} \right),
\nonumber \\
 I_{(Q)X}^Y &\equiv& \tQ^{ i}_X\Qt_i^{Y}=\left(
\begin{array}{cc}
-\frac{\delta_a^b\chi^2-\chi_a\chi^b}{1-\chi^2} &
-\frac{ {\eps_a}^{bc}\chi_c}{1-\chi^2} \\
-\frac{ {\eps_a}^{bc}\chi_c}{1-\chi^2} &
\frac{\delta_a^b-\chi_a\chi^b}{1-\chi^2}
\end{array} \right). \label{Qi-Qi}
\eeq
Despite their complicated form the relations (\ref{Qi-Qi}) have a simple
interpretation.
The matrices $I_{(P)X}^Y$ and $I_{(Q)X}^Y$ are
projectors on $\tP$ and $\tQ$-multiplets in the linear space spanned by
these vectors. Indeed, the following equalities are fulfilled due to
the second class constraints $\phi^{ij}$:
\beq
& I_{(P)Z}^Y I_{(P)X}^Z= I_{(P)X}^Y, \qquad
I_{(Q)Z}^Y I_{(Q)X}^Z= I_{(Q)X}^Y, & \nonumber\\
& I_{(P)X}^Y +I_{(Q)X}^Y=\delta_X^Y, & \nonumber \\
&I_{(P)X}^Y \tP^i_Y=\tP^i_X, \quad I_{(P)X}^Y \Pt_i^X=\Pt_i^Y,
\qquad
I_{(Q)X}^Y \tQ^i_Y=\tQ^i_X, \quad I_{(Q)X}^Y \Qt_i^X=\Qt_i^Y,&
\label{propI} \\
&I_{(P)X}^Y \tQ^i_Y=0, \quad I_{(P)X}^Y \Qt_i^X=0, \qquad
I_{(Q)X}^Y \tP^i_Y=0, \quad I_{(Q)X}^Y \Pt_i^X=0.& \nonumber
\eeq

As a result one can check that
\be
(D_1^{-1})_{(kl)(mn)}=\frac{1}{8}\left(1+\frac{1}{\beta^2}\right)
\left( (\Qt\Qt)_{kl}(\Qt\Qt)_{mn}-(\Qt\Qt)_{km}(\Qt\Qt)_{ln}-
(\Qt\Qt)_{kn}(\Qt\Qt)_{lm} \right)
\ee
gives $D_1^{(ij)(kl)}(D_1^{-1})_{(kl)(mn)}=\delta^{\{i}_{\{m}
\delta^{j\}}_{n\}}$.
The Dirac brackets of the canonical variables take the form:
\beq
\{ \tPb_X^i,\tPb_Y^j\}_D&=&0, \nonumber \\
\{ A^X_i,\tPb_Y^j\}_D&=&\delta_i^j\delta^X_Y-\frac12\left(g^{XZ}-
\frac{1}{\beta}\Pi^{XZ}\right)\left(\tQ^j_Z\Qt_i^W+\delta^j_i I_{(Q)Z}^W
\right)g_{WY}, \label{comm} \\
\{ A^X_i,A^Y_j\}_D&=&-\{A_i^X,\phi^{kl}\}(D_1^{-1})_{(kl)(mn)}
\{\psi^{mn},A^Z_r\}\{\tPb^r_Z,A_j^Y\}_D \nonumber \\
&&-\{A_i^X,\tPb^r_Z\}_D\{A_r^Z,\psi^{mn}\}(D_1^{-1})_{(mn)(kl)}
\{\phi^{kl},A^Y_j\}.  \nonumber
\eeq
The Dirac brackets (\ref{comm}) represent the commutation algebra
which should be used at the quantum level.

\section{Notes on the Immirzi parameter problem}

\subsection{Path integral quantization}

In this section we are going to compare the path integrals
constructed for the formulations with different values of the Immirzi
parameter.

Consider the path integral for the theory with the action (\ref{Sd3}).
Choose the gauge fixing condition in the following form:
\be
f^{\alpha}(\tP,A)+g^{\alpha}({\cal N)}=0.
\label{gauge} \ee
Here ${\cal N}^{\alpha}=({\cal N}_{\G}^X,\nd^i,\tNn)$ is the set of
the Lagrange multipliers.
Let us restrict ourselves to a definite class of gauges, which are
the Yang--Mills (YM) gauges introduced in the work \cite{AV}.
They are described by the gauge fixing functions (\ref{gauge}) with
two additional conditions: (a) $g^{\alpha}$ don't depend on
${\cal N}_{\G}^X$, (b) $f^{\alpha}$ don't depend on $A_i^X$.
Thus we are not allowed to fix the Lagrange multipliers
for the Gauss constraint and impose gauge conditions
on the connection.
Due to these restrictions the multighost interaction terms do not appear
in the effective action and the path integral is given by
the ordinary phase space path integral \cite{Faddeev}:
\beq
Z[\bar j,\bar J] &=&\int {\cal D}A_i^X {\cal D}\tPb^i_X
{\cal D}{\cal N}^{\alpha}  {\cal D}c^{\alpha}
{\cal D}\bar c_{\alpha}
\sqrt{|\Delta|}
\delta (\phi^{ij}) \delta (\psi^{ij})
\delta (f^{\alpha} {+} g^{\alpha})
\nonumber \\
&\times &
\exp\left[ i\int dt\,
(L_{eff}'+ j_a^iA_i^a+ J_i^a\tP^i_a)\right],
\label{pi} \eeq
where
\be L_{eff}'= L_{\beta}-
i\bar c_{\beta}\Bigl(
\frac{\partial g^{\beta}}{\partial {\cal N}^{\alpha}}\partial_t -
\frac{\partial g^{\beta}}{\partial {\cal N}^{\gamma}}
C^{\gamma}_{\alpha \delta}{\cal N}^{\delta}+
\left\{ \Phi^{(\beta)}_{\alpha}, f^{\beta} (\tP) \right\}^{(\beta)}_D
\Bigr) c^{\alpha}. \label{Leff}
\ee
$\Delta$ is taken from (\ref{Delta}). We introduced the sources $J$
for the fields $\tP$ rather than $\tPb$ to simplify the comparison
of the path integrals for different $\beta$'s. It does not change
their sense, since $\tPb$ is expressed unambiguously through $\tP$
due to (\ref{p-q}), (\ref{pb-q}).
In addition, all
physical operators (as, e.g., the area operator)
should not depend on $\beta$ and so they are
expressed more naturally through $\tP$.

Let us investigate the dependence of the path integral (\ref{pi}) on
the Immirzi parameter $\beta$.
We shall try to rewrite the path integral (\ref{pi}) in terms of variables
corresponding to $\beta=\infty$
($\tP^{(\infty)}=\tP$) and thus independent on
the parameter. For this each $\beta$-dependent contribution will be
extracted and discussed.
There are several sources of such contributions.

The first source is the delta function $\delta({\cal G}^{(\beta)}_X)$
appearing after integration over ${\cal N}_{\G}^X$.
(We can perform this integration due to the first condition on
the YM gauge.)
Since ${\cal G}^{(\beta)}_X=-\Lambda_X^Y\Pi_Y^Z{\cal G}^{(\infty)}_Z$,
it gives the multiplier
\be m_1={\rm det}^{-1}(\Lambda\Pi) =
\prod_{x,t}\left(1+\frac{1}{\beta^2}\right)^{-3}. \label{m1} \ee

The second place where $\beta$-dependent terms arise is the action.
However, in \cite{AV} it was established
that the imaginary part of the Ashtekar action (\ref{Asht})
vanishes on the surface
of the second class constraints and the Gauss and Lorentz constraints.
But it is just that part of the action (\ref{Sd1}), which introduces
the $\beta$-dependence, what can be seen from (\ref{Sd2}).
In addition, the path integral (\ref{pi})
contains the delta functions of
the second class constraints $\phi^{ij}$ and $\psi^{ij}$ as well as
the delta functions of the constraint ${\cal G}^{(\beta)}_X$.
Thus $S_{\beta}$ can be reduced to ${\rm Re}S_A=S_{(\infty)}$
and does not introduce any $\beta$-dependence.

The measure gives three contributions. The first one comes from
${\cal D}\tPb^i_X={\rm det}^3\left(\Lambda\Pi\right){\cal D}\tP^i_X$. Thus
\be m_2={\rm det}^{3}(\Lambda\Pi) =
\prod_{x,t}\left(1+\frac{1}{\beta^2}\right)^{9}. \label{m2}\ee
The second contribution is given by $\sqrt{|\Delta|}={\rm det} (D_1)$.
Since $\beta$ enters in (\ref{D1}) in the common multiplier only
we obtain
\be m_3={\rm det}\left(\frac{\beta^2}{1+\beta^2} I_{6\times 6}\right) =
\prod_{x,t}\left(1+\frac{1}{\beta^2}\right)^{-6}. \label{m3} \ee
Finally, the last contribution can arise from the Faddeev-Popov determinant.
However, it turns out that the determinants in the formulations with
different $\beta$'s coincide. The reason for this is
the $\beta$-independence of the structure constants of
the constraint algebra (\ref{algA})
and the remarkable relations (\ref{remark}) between the Dirac brackets.
Since the YM gauge does not allow for $f^{\alpha}$ to depend on
the connection, one can replace the bracket in (\ref{Leff})
by the bracket independent on $\beta$:
$\bigl\{ \Phi^{(\infty)}_{\alpha}, f^{\beta} (\tP) \bigr\}^{(\infty)}_D.$
Thus the Immirzi parameter does not appear in
the second term of the effective action (\ref{Leff})
which produces the Faddeev-Popov determinant.

As a result the dependence of the path integral on $\beta$
is contained in three
multipliers (\ref{m1}), (\ref{m2}) and (\ref{m3}) which cancel each other.
Thus we arrive to the conclusion that at least at the formal level
the path integral for quantum gravity
is independent on the Immirzi parameter.

\subsection{Loop approach}

Relying on the formalism developed in section 3 one can try to work out
an alternative loop approach.
The key point is that although $A_i^X$ is noncommutative due to
(\ref{comm}) it is transformed as a true connection under the gauge
transformations (\ref{trans}).
Thus the Wilson loop operator can be constructed
\be
U_{\alpha}(a,b)={\cal P}\exp\left(\int_a^b dx^i A_i^X T_X\right), \label{hol}
\ee
where $\alpha$ is a path between two points $a$ and $b$,
$T_X$ is a gauge generator.
Using these operators one can try to construct
the full Hilbert space of quantum gravity in the same way as
it is done in the standard loop approach.
However, we encounter the serious obstacle on this way since
the simple canonical commutation relations are changed now by
the complicated commutators (\ref{comm}). Due to this the operators
like (\ref{hol}) fail to form the loop algebra. Whether there is
an another algebra with an explicit geometric interpretation, which
substitutes the loop algebra, is the crucial question
for the formalism.
Only the existence of this algebra will provide
a solid ground for these speculations.
So far we have not been able to do it because of a very complicated
structure of the last relation in (\ref{comm}).

Even if such an algebra is found
there are two most serious difficulties arising owing to the new
commutation relations.
The first one is connected with the non-compactness
of the Lorentz group, which is suggested
to be used to define a scalar product in this approach as SU(2) does.
However, with the other hand it can open a possibility to tackle
the problem of time in canonical quantum gravity.
Also a necessity to deal with a non-compact gauge group is stressed
in the paper \cite{Sam}.

The second difficulty is
that the connection representation is not
applicable in this framework due to the noncommutativity of $A_i^X$.
Nevertheless one may hope to extract some
physical results relying on algebraic relations only
or even to develop the $\tE$-representation.
For example, one can try to obtain the spectrum of the area operator
\be A_S=\int_S d^2s\sqrt{n_in_jg^{ij}}\ee
from its commutators with the Wilson loops,
if the vacuum state is an eigenstate of the area operator.
Here the metric $g^{ij}$ is taken from (\ref{3-met}).
These commutators should be calculated
using the quantum version of the commutation relations (\ref{comm}).

In this connection we can observe that
\be \{ A^X_k , g^{ij}\}_D =
g^{XY}(\delta_k^i \tP^j_V+\delta_k^j \tP^i_V),\ee
i.e., the additional contribution of the Dirac bracket
cancels the dependence on $\beta$. It is not clear how this fact reflects
in the spectrum, but it shows in what way the Immirzi parameter can
disappear from it.
Remind, however, that all this will have a sense only after a substitute for
the loop algebra is found.

It is worth to notice the crucial difference between the
covariant formalism, which is outlined here,
and the conventional loop approach with
the Ashtekar--Barbero gravity in the ground.
In the covariant formalism we have a unique connection for all values of
$\beta$ instead of a one-parameter family of connections in
the Ashtekar--Barbero gravity. Since we did not solve the second class
constraints, nothing can be added to $A_i^X$ to obtain
a true so(4,C)-connection
\footnote{
Indeed one can construct several quantities from $\tQ$ and $\Qt$ multiplets
which are transformed either homogeneously or as so(4,C)-connection.
They are similar to the Christoffel connection. So one could use
them to obtain a family of so(4,C)-connections. However, it turns out
that all of them are transformed in a "wrong" way under the diffeomorphism
constraint. Thus there is only one connection with all right
transformation lows.
}.

This provides us with a new look at the generalized
Wick rotation connecting the formulations with different $\beta$'s
\cite{Wick}.
From (\ref{multHP}) and (\ref{Sd3}) we observe that in
our approach they differ by a shift of
the dynamical triad multiplet (and may be a choice of the basis
of the adjoint representation). This moves the accent from the connection
onto the triad. Let us remind that in the Barbero approach
based on the SU(2) gauge subgroup
the Wick transformation changes the connection rather than the triad.
In our case the connection is unique but there are two triad multiplets.
Just this allows to form a one-parameter family of triad multiplets
rather than connections.
However, we have not succeeded so far in representing this shift
by a canonical operator.

\section{Conclusions}

In this paper we have suggested a new hamiltonian formulation
of general relativity based on the full SO(3,1) gauge group.
Without any preliminary gauge fixing we have constructed
the hamiltonian formalism for the generalized Hilbert-Palatini gravity
which encompasses the Ashtekar--Barbero gravity \cite{barb}.
It turns out to be covariant under so(4,C) transformations, and
the set of the canonical variables forms multiplets in
the adjoint representation of this algebra.

Then the developed formalism has been applied to the investigation
of dependence of the quantum gravity on the Immirzi parameter.
It has been shown in the framework of the formal path integral quantization
that the formulations with different values of $\beta$
should be all equivalent. We also speculate on possible extension of
the loop approach to the theory developed above without giving, however,
any rigorous results.
To be able to develop a loop quantization of the suggested formalism
one should overcome a number of difficulties.
The main one is to find a loop representation of the algebra
of the obtained Dirac brackets.
It is quite nontrivial due to the noncommutativity of the connection
and since the inverse triad multiplets are involved.
It is not obvious whether this is possible at all.
But correct transformation properties with respect to the full
Lorentz group of the connection entering
the Wilson loop operator give the hope that a success may be achieved
in this way.

The noncommutativity of the connection is the main technical difficulty of
the approach. But at the same time it may indicate
the appearance of the noncommutative geometry in the framework.
However, there is no a direct analogy between the observed noncommutativity
and the one arising in the modern superstring theories, for instance.
The main difference is that the former appears already at the classical
level.
Nevertheless it would be interesting to see how the methods of
the noncommutative geometry, if it actually appears here, work in canonical
quantum gravity.

Finally, the new realization
of the generalized Wick rotation suggested in the end of section 4 allows
to review the question how this transformation is implemented in
the quantum theory \cite{Rov-Tim}.
It is very desirable to understand how the fact that this transformation
is not canonical even at the classical level combines with
the equivalence of the formulations with different $\beta$'s established
using path integral.

\section*{Acknowledgements}

 The author is very grateful to D.V.Vassilevich for
helpful and valuable discussions.
This work was supported by Young
Investigator Program.

\appendix

\section{Matrix algebra}

Introduce the matrices connecting different triad
multiplets:
\beq
\tP^i_X=\Pi_X^Y\tQ^i_Y,&\qquad &
\Pi^{Y}_X =\left(
\begin{array}{cc}
0&1 \\ -1&0
\end{array}
\right)\delta_a^b,   \label{p-q}     \\
\tPb^i_X=\Lambda_X^Y\tQ^i_Y,&\qquad &
\Lambda^{Y}_X =\left(
\begin{array}{cc}
-\frac{1}{\beta}&1
\\ -1& -\frac{1}{\beta}
\end{array}
\right)\delta_a^b.     \label{pb-q}
\eeq
Although the matrices $\Pi_X^Y$ and $\Lambda_X^Y$ do not possess
any symmetry the matrices $\Pi^{XY}=g^{XZ}\Pi_Z^Y$ and
$\Lambda^{XY}=g^{XZ}\Lambda_Z^Y$ turn out to be symmetric.
Besides, the following relations are fulfilled:
\beq
&\Lambda^Y_X=\Pi^Y_X-\frac{1}{\beta}\delta^Y_X,& \\
& (\Pi^{-1})^Y_X=-\Pi^Y_X, \quad (\Lambda^{-1})^Y_X=-\frac{\Lambda^Y_X
+\frac{2}{\beta} \delta^Y_X}{1+\frac{1}{\beta^2}}=
-\frac{\Pi^Y_X+\frac{1}{\beta} \delta^Y_X}{1+\frac{1}{\beta^2}}.&
\label{inverse} \eeq
Due to these relations $\Pi$, $\Lambda$ and their inverse commute
with each other.
Furthermore, they commute with the structure constants
in the following sense:
\be f^{XYZ'}\Pi_{Z'}^Z=f^{XY'Z}\Pi_{Y'}^Y. \label{com-f} \ee
One more useful relation is
\be
f^T_{XY} f^W_{TZ}=-g_{XZ}\delta^W_Y+g_{YZ}\delta^W_X+
\Pi_{XZ}\Pi^W_Y-\Pi_{YZ}\Pi^W_X.  \label{ff}
\ee
Being established in the basis (\ref{multHP}), all these relations
are valid in an arbitrary basis.

\section{Dual representation}

There is a special choice of the basis of the adjoint representation
of so(4,C)
which is closely connected with the variables used in Barbero's formulation
\cite{barb}.
Let us express the action (\ref{Sd1}) in terms of
a connection reduced in the "time" gauge to the Barbero connection
and without the star operator. However, in contrast to the self-dual case
it can be done only using two connections. Since there is no
way to decide which connection should contain $\beta$, we define them
in the symmetric way:
\be A_{(1)}^{\alpha\beta}=\frac12(\omega^{\alpha\beta}-\ba\star
\omega^{\alpha\beta}),\qquad
A_{(2)}^{\alpha\beta}=\frac12(\omega^{\alpha\beta}-\bb\star
\omega^{\alpha\beta}). \ee
The field strength two-forms are
\beq \F_{(1)}^{\alpha\beta}&=&dA_{(1)}^{\alpha\beta}+
A^{\alpha}_{(1)\gamma}\wedge
A_{(1)}^{\gamma\beta}=\frac12(\Omega^{\alpha\beta}-\ba\star\Omega^{\alpha
\beta})-\frac14(1+\ba^2)\omega^{\alpha}_{\gamma}\wedge\omega^{\gamma
\beta}, \label{F} \\
\F_{(2)}^{\alpha\beta}&=&dA_{(2)}^{\alpha\beta}+
A^{\alpha}_{(2)\gamma}\wedge
A_{(2)}^{\gamma\beta}=\frac12(\Omega^{\alpha\beta}-\bb\star\Omega^{\alpha
\beta})-\frac14(1+\bb^2)\omega^{\alpha}_{\gamma}\wedge\omega^{\gamma
\beta}. \label{G} \eeq
They obey the relation
\be \frac{1}{\ba^2-\bb^2}\left( (1+\bb^2)\F_{(1)}^{\alpha\beta}-
(1+\ba^2)\F_{(2)}^{\alpha\beta}\right)=-\frac{1}{2}\Omega^{\alpha\beta}-
\frac{1-\ba\bb}{2(\ba+\bb)}\star\Omega^{\alpha\beta}. \label{kriv} \ee
It means that we can rewrite (\ref{Sd1}) in the required form if we
set $\beta=\frac{\ba+\bb}{1-\ba\bb}$:
\be
S_{(\beta)}=-\frac{1}{\ba^2-\bb^2}
\int \eps_{\alpha\beta\gamma\delta}e^{\gamma}\wedge e^{\delta}
\wedge((1+\bb^2)\F_{(1)}^{\alpha\beta}-(1+\ba^2)\F_{(2)}^{\alpha\beta}).
\label{Sd4}\ee

The action $S_{(\beta)}$ is invariant under the
"duality" transformation
$A_{(1)}\longleftrightarrow A_{(2)},\ \ba\longleftrightarrow \bb$.
It is just a generalization of the selfduality leading to
the Ashtekar action,
which can be obtained from $S_{(\beta)}$ in the limit $\ba=i$ (or $\bb=i$).
Another useful limit is $\bb=0,\ba\longrightarrow \infty$,
which together with the redefinition of the connection
$A_{(1)}\longrightarrow
\frac{1}{\ba}A_{(1)}$ leads to the Hilbert--Palatini gravity.

Of course, the observed duality is trivial in the sense it is only
a change of variables. Besides, indeed the theory depends on
one parameter only. So one of $\beta$'s can be fixed in a way.
For example, we can set $\bb=0$. But we shall keep them to be arbitrary
to retain the duality.

The 3+1 decomposition of the action (\ref{Sd4}) can be obtained
using the definitions (\ref{tetrad}). The result is represented
in the covariant form (\ref{Sd3}), but the natural choice of the multiplets
looks as
\beq
\G_X&=&
\Bigl(\sqrt{\frac{1+\bb^2}{\bb^2-\ba^2}} \G^{(1)}_a, \
\sqrt{\frac{1+\ba^2}{\ba^2-\bb^2}}\G^{(2)}_a\Bigr), \nonumber \\
A^X_i&=&
\Bigl(\sqrt{\frac{1+\bb^2}{\bb^2-\ba^2}} A^{(1)a}_i,
\sqrt{\frac{1+\ba^2}{\ba^2-\bb^2}}A_i^{(2)a}\Bigr),  \\
\tPb_X^i&=&
\Bigl(\sqrt{\frac{1+\bb^2}{\bb^2-\ba^2}} P^i_{(1)a}, \
\sqrt{\frac{1+\ba^2}{\ba^2-\bb^2}}P_{(2)a}^i\Bigr), \nonumber
\eeq
where we introduced the following fields:
\beq
&  P^i_{(1)a}={\eps_a}^{bc}\tE^i_b\chi_c+\ba\tE^i_a, \qquad
  P^i_{(2)a}={\eps_a}^{bc}\tE^i_b\chi_c+\bb\tE^i_a,& \label{PQ}\\
&  A^{(1)a}_i=\xi_i^a-\ba\zeta_i^a, \qquad
  A^{(2)a}_i=\xi_i^a-\bb\zeta_i^a, & \label{AB} \\
& \G^{(1)}_a=\partial_i P^i_{(1)a} -{\eps_{ab}}^c A^{(1)b}_i P^i_{(1)c}+
\frac{1+\ba^2}{(\ba-\bb)^2}{\eps_{ab}}^c(A^{(1)b}_i-A^{(2)b}_i)
(P^i_{(1)c}-P^i_{(2)c}),& \label{GA}\\
 & \G^{(2)}_a=\partial_i P^i_{(2)a} -{\eps_{ab}}^c A^{(2)b}_i P^i_{(2)c}+
\frac{1+\bb^2}{(\ba-\bb)^2}{\eps_{ab}}^c(A^{(1)b}_i-A^{(2)b}_i)
(P^i_{(1)c}-P^i_{(2)c}).& \label{GB}
\eeq
These multiplets are related with the multiplets (\ref{multHP}) by
the following matrix:
\beq
 U_X^Y&=&\left(
\begin{array}{cc}
-\ba\sqr{\frac{1+\bb^2}{\bb^2-\ba^2}} &
-\bb\sqr{\frac{1+\ba^2}{\ba^2-\bb^2}}  \\
\sqr{\frac{1+\bb^2}{\bb^2-\ba^2}} &
\sqr{\frac{1+\ba^2}{\ba^2-\bb^2}}
\end{array}
\right)\delta_a^b.
 \eeq
In this new basis the structure constants become more complicated.
They are obtained using (\ref{GA}) and (\ref{GB}):
\be
\begin{array}{cc}
f_{A_1 A_2}^{A_3}=\frac{1+2\ba\bb-\bb^2}{(\ba-\bb)^2}
\sqr{\frac{\bb^2-\ba^2}{1+\bb^2}}\eps^{A_1 A_2 A_3},&
f_{B_1 B_2}^{B_3}=\frac{1+2\ba\bb-\ba^2}{(\ba-\bb)^2}
\sqr{\frac{\ba^2-\bb^2}{1+\ba^2}}\eps^{B_1 B_2 B_3},\\
f_{A_1 B_2}^{A_3}=-\frac{\sqr{(1+\ba^2)(\ba^2-\bb^2)}}
{(\ba-\bb)^2}\eps^{A_1 B_2 A_3},&
 f_{A_1 B_2}^{B_3}=-\frac{\sqr{(1+\bb^2)(\bb^2-\ba^2)}}
{(\ba-\bb)^2}\eps^{A_1 B_2 B_3},\\
f_{B_1 B_2}^{A_3}=-\frac{\sqr{(1+\bb^2)(\bb^2-\ba^2)}}
{(\ba-\bb)^2}\eps^{B_1 B_2 A_3}, &
f_{A_1 A_2}^{B_3}=-\frac{\sqr{(1+\ba^2)(\ba^2-\bb^2)}}
{(\ba-\bb)^2}\eps^{A_1 A_2 B_3}.
\end{array}  \label{alg2}
\ee

Notice, that in this basis the hamiltonian constraint can be expressed
through the dynamical multiplet only
\be
H=\frac12 \tPb^i_X \tPb^j_Y f_{XY}^Z F_{ij}^Z. \label{ham2}
\ee
As it is easy to see the indices in (\ref{ham2})
are contracted with help of
the unit matrix. It gives an invariant expression due to the existence
of the unit invariant form. This is a consequence of the full
antisymmetry of the structure constants (\ref{alg2}).
However, such representation is not covariant under change of the basis.

\section{Constraint algebra}
Define the smeared constraints:
\begin{eqnarray}
&&{\cal G}(n)=\int d^3x\, n^X{\cal G}_X, \qquad
H(\tN )=\int d^3x\, \tN H,
\nonumber \\
&&{\cal D}(\vec N)=\int d^3x\, N^i(H_i+A_i^X{\cal G}_X).
\end{eqnarray}
They obey the following algebra:
\begin{eqnarray}
&&\left\{ {\cal G}(n) ,{\cal G}(m) \right\}_D={\cal G}(n\times m),
\nonumber \\
&&\left\{ {\cal D}(\vec N) ,{\cal D}(\vec M) \right\}_D=
-{\cal D}([\vec N ,\vec M ]),\nonumber \\
&&\left\{ {\cal D}(\vec N) ,{\cal G}(n) \right\}_D=-
{\cal G}( N^i\partial_in), \nonumber \\
&&\Bigl\{ H(\tN ) ,{\cal G}(n) \Bigr\}_D =0, \label{algA} \\
&&\Bigl\{ {\cal D}(\vec N) ,H(\tN ) \Bigr\}_D=
-H({\cal L}_{\vec N}\tN ), \nonumber \\
&&\Bigl\{ H(\tN ),H(\tM ) \Bigr\}_D =
{\cal D}(\vec K)-{\cal G}(K^jA_j), \nonumber
 \end{eqnarray}
where
\begin{eqnarray}
&&(n\times m)^X=f^X_{YZ}n^Ym^Z,\qquad
{\cal L}_{\vec N}\tN =
N^i\partial_i \tN-\tN\partial_iN^i, \nonumber \\
&&[\vec N ,\vec M ]^i=
N^k\partial_kM^i-M^k\partial_kN_i, \label{not1} \\
&& K^j=(\tN\partial_i\tM-\tM\partial_i\tN)\tQ^i_X\tQ^j_Y g^{XY}.
\nonumber
\end{eqnarray}

\end{document}